\documentclass[prb,twocolumn,aps,superscriptaddress,longbibliography]{revtex4-2}
\usepackage{graphicx}
\usepackage{dcolumn}
\usepackage{bm}
\usepackage{amssymb}
\usepackage{amsmath}
\usepackage{subfigure}
\usepackage{physics}
\usepackage[binary-units]{siunitx}
\usepackage{sublabel}
\usepackage[utf8]{inputenc}

\usepackage{hyperref}
\usepackage{floatrow}
\usepackage{graphicx}
\usepackage{xcolor}
\usepackage[normalem]{ulem}

\usepackage{bbm}
\hypersetup{colorlinks = true, linkcolor=blue, citecolor=blue, urlcolor=}
\usepackage[T1]{fontenc}

\newcommand{\SPHIDE}[1]{{}}

\begin{document}


\title{Decoupling Nuclear Spins via Interaction-Induced Freezing in Nitrogen Vacancy Centers in Diamond}

\author{Abhishek Kejriwal}
\thanks{These authors contributed equally to this work.}
\affiliation{Department of Physics, Indian Institute of Technology Bombay, Powai, Mumbai-400076, India}

\author{Dasika Shishir}
\thanks{These authors contributed equally to this work.}
\affiliation{Department of Electrical Engineering, Indian Institute of Technology Bombay, Powai, Mumbai-400076, India}


\author{Sumiran Pujari}
\affiliation{Department of Physics, Indian Institute of Technology Bombay, Powai, Mumbai-400076, India}

\author{Kasturi Saha}
\thanks{kasturis@ee.iitb.ac.in}
\affiliation{Department of Electrical Engineering, Indian Institute of Technology Bombay, Powai, Mumbai-400076, India}

\date{\today}
\begin{abstract}

Nitrogen-Vacancy (NV) centers in diamonds provide a room-temperature platform for various emerging quantum technologies, e.g. the long nuclear spin coherence times as potential quantum memory registers. We demonstrate a freezing protocol for an NV center to isolate its intrinsic nuclear spin from a noisy electromagnetic environment.
Any initial state of the nuclear spin can be frozen when the hyperfine-coupled electron and nuclear spins are simultaneously driven with unequal Rabi frequencies. 
Through numerical simulations, we show that our protocol can effectively shield the nuclear spin from strong drive or noise fields. We also observe a clear suppression of quantum correlations in the frozen nuclear spin regime by measuring the quantum discord of the electron-nuclear spin system. These features can be instrumental in extending the storage times of NV nuclear-spin based quantum memories in hybrid quantum systems.

\end{abstract}

\maketitle


\section{Introduction}\label{sec1}

Nitrogen-Vacancy (NV) centers in diamond \cite{DOHERTY20131, Balasubramanian2009} have emerged as a leading platform for quantum technology demonstrations with applications ranging from precision quantum metrology \cite{2008,Maze2008,2010,McGuinness2011} to quantum computation \cite{Wrachtrup_2006,doi:10.1126/science.1157233}. NV centers are attractive candidates for these applications by virtue of their features such as easy optical initialization and readout scheme \cite{PhysRevLett.92.076401}, fast coherent manipulation using electromagnetic fields \cite{doi:10.1126/science.1181193}, and high coherence times under ambient conditions \cite{Balasubramanian2009,PhysRevB.86.045214}. Another significant feature that expands the range of NV center-based applications is the ability to initialize and readout proximal nuclear spins by leveraging their hyperfine coupling to the NV electron spin \cite{doi:10.1126/science.1131871,Gaebel2006,PhysRevLett.97.087601,PhysRevLett.93.130501,doi:10.1126/science.1189075}. \\

The nuclear spins associated with the NV center offer unique prospects as room temperature quantum registers, with reasonably long coherence times, and thus are potential candidates for quantum memory applications. Thus, extending the nuclear spin lifetimes and effectively shielding the nuclear spin from environmental noise is an important problem in quantum information science. 
Recent work based on Rydberg atoms~\cite{RevModPhys.82.2313,2012_rydberg} has demonstrated a freezing effect \cite{2019_rydberg}, whereby the state dynamics of a Rydberg atom freezes by a suitable application of external control fields to the strongly-interacting Rydberg atoms. 
Such interaction-induced freezing effects have caught common interest due to their applications, such as local spin control~\cite{PhysRevA.81.040303,PhysRevLett.113.020506}. This motivates leveraging the hyperfine interaction between the NV center's electron and nuclear spins to develop a protocol for nuclear spin freezing by exploiting the natural separation of the gyromagnetic scales of the electron and nuclear spins of the NV system.  This will allow the freezing of the NV center's nuclear spin dynamics to its initial state through the simultaneous driving of the electron spin and nuclear spin using a protocol tailored to the NV system whose details will be specified in the main text. The electron spin remains dynamic during this process, which amounts to \emph{effectively} decoupling the electron and nuclear spin dynamics. It turns out that doing the inverse (i.e., freezing the electron spin), even though possible, requires unrealistic values for the drive fields and thus is not the focus here. \\

\begin{figure}[!t]
	\centering
	\subfigure[]{\includegraphics[width=0.4\textwidth]{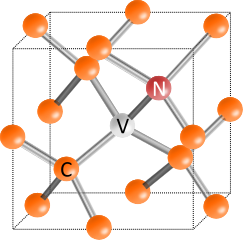}\label{1a}}
	\quad
	\subfigure[]{\includegraphics[width=0.35\textwidth]{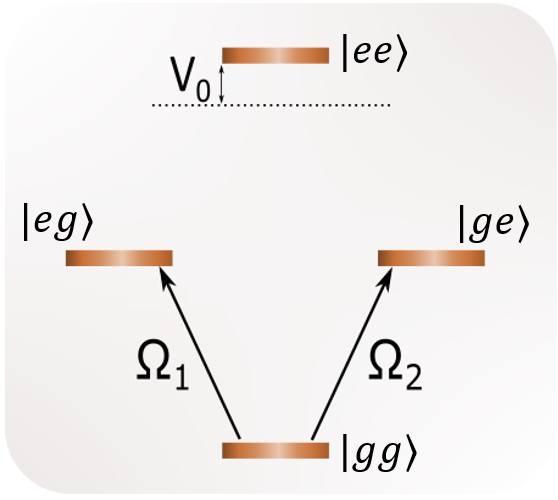}\label{1b}}
	\quad
	\subfigure[]{\includegraphics[width=0.8\textwidth]{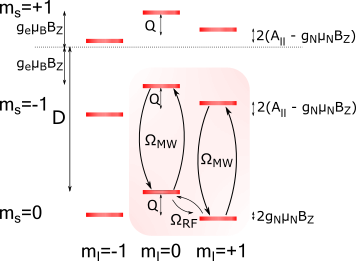}\label{1c}}
	\quad
	
	\caption{(a) Physical structure of NV center in a diamond lattice. (b) Energy diagram of two interacting spins with interaction potential $V_{0}$, driven by resonant fields with Rabi frequencies $\Omega_{1}$ and $\Omega_{2}$.  (c) NV center's electron and nuclear energy level diagram in presence of hyperfine coupling of the NV electron spin with NV center's intrinsic \textsuperscript{14}N nuclear spin. A microwave field is used to drive the electronic transition, whereas an RF field is used to drive the nuclear spin transition.  }
	\label{fig:1}
\end{figure}

\begin{figure*}[!tbp]
	\centering
	\subfigure[]{\includegraphics[width=0.23\textwidth]{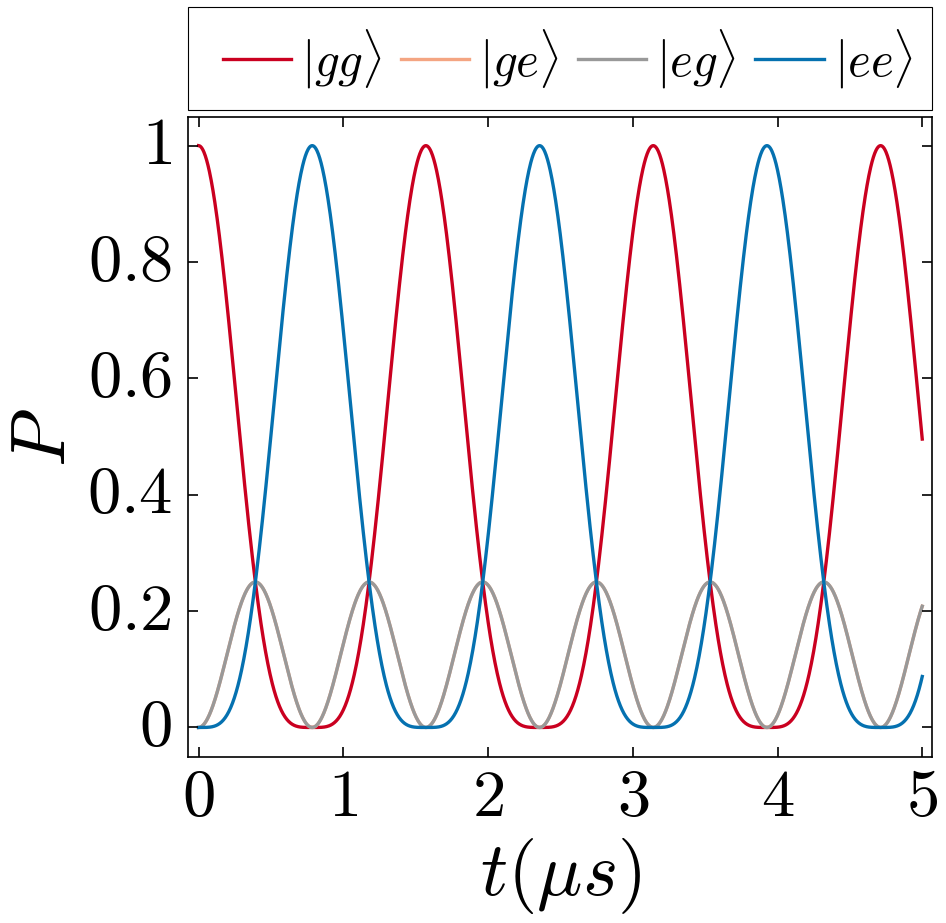}\label{2a}}
	\quad	
	\subfigure[]{\includegraphics[width=0.23\textwidth]{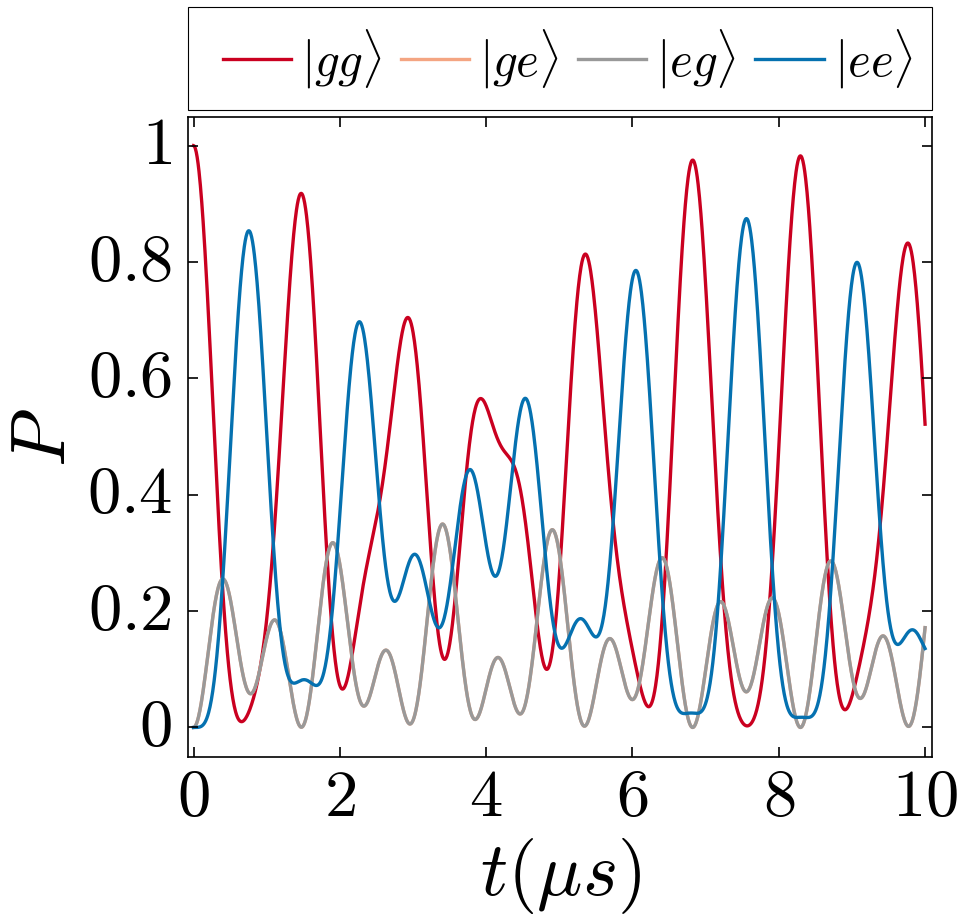}\label{2b}}
	\quad
	\subfigure[]{\includegraphics[width=0.23\textwidth]{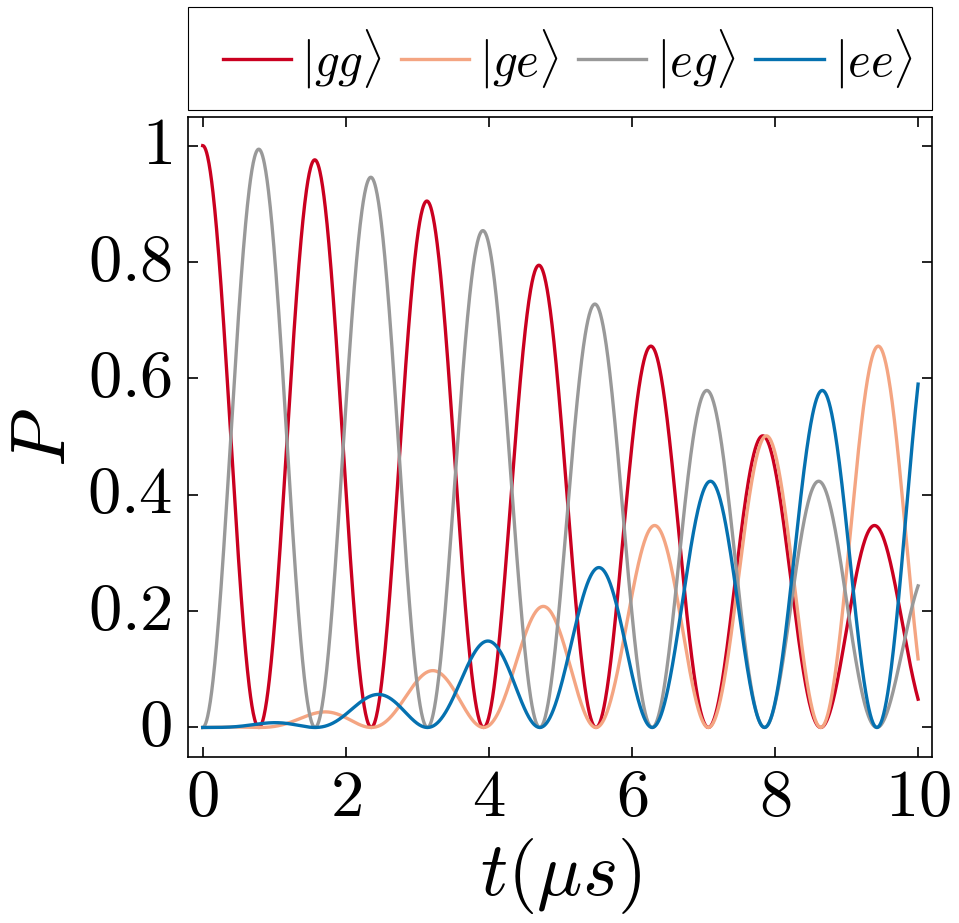}\label{2c}}
	\quad
	\subfigure[]{\includegraphics[width=0.23\textwidth]{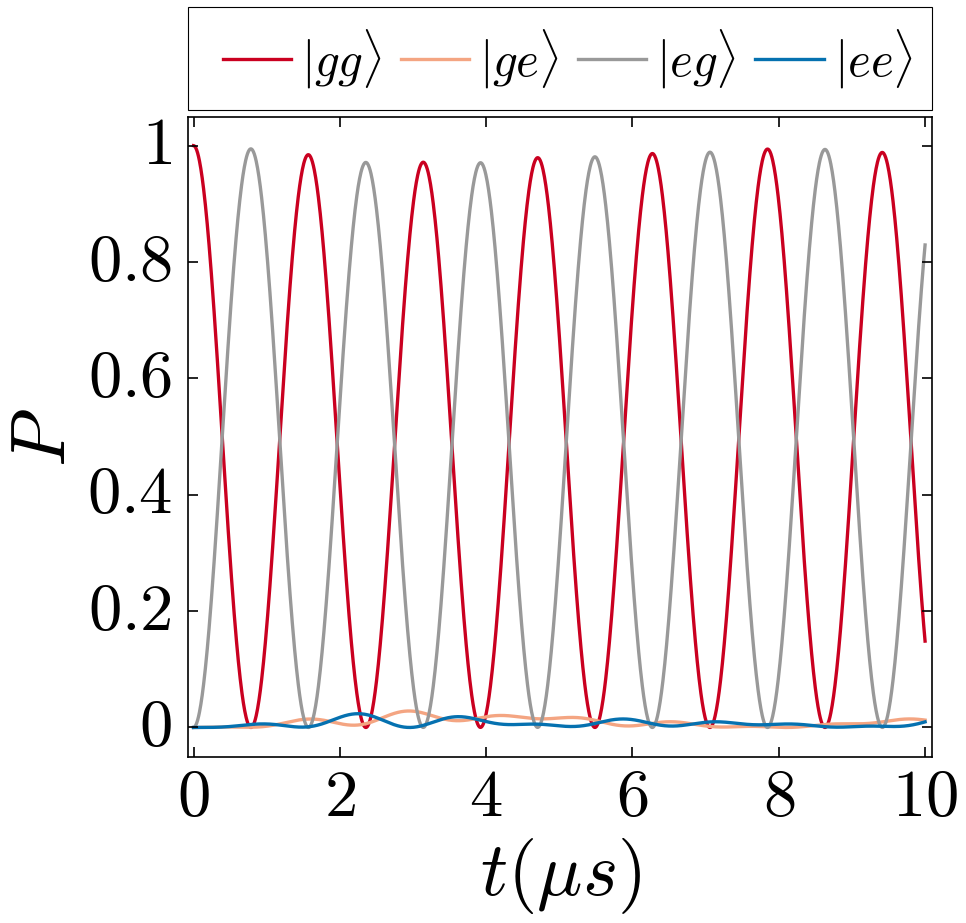}\label{2d}}
	\quad	
	\caption{Plots for state dynamics of two spins, interacting with potential $V_0$, and driven simultaneously and independently by drive fields with Rabi frequencies $\Omega_1$ and $\Omega_2$ respectively. (a) State dynamics for spins placed far from each other and driven by fields with $\Omega_{1}=\Omega_{2}$ ( $\Omega_{1}=2$MHz, $\Omega_{2}=2$MHz, $V_0=0$). Both spins oscillate between their ground and excited states and the system thus oscillates between $|gg\rangle$ and $|ee\rangle$. (b) State dynamics for spins placed close to each other such that $V_{0}$ is a high non-zero value and both spins driven at the same Rabi frequency  ($\Omega_{1}= \Omega_{2}= 2$MHz, $V_{0}= 2$ MHz).  Excited states of both the spins get populated during the dynamics and system oscillates between all the four basis states. (c) State dynamics for spins placed far from each other and driven by fields with $\Omega_{1}>>\Omega_{2}$ ( $\Omega_{1}=2$MHz, $\Omega_{2}=100$kHz, $V_{0}=0$).  All four states of the two-qubit system get populated during the dynamics. (d) State dynamics for spins placed close to each other and driven by highly different Rabi frequencies such that $V_{0}, \Omega_{1}>>\Omega_{2}$ ( $\Omega_{1}=2$MHz, $\Omega_{2}=100$kHz, $V_0=2$MHz).  The system oscillates only between $|gg\rangle$ and $|eg\rangle$ and the population of states $|ge\rangle$ and $|ee\rangle$ stay fixed at $0$. High bias in the drive fields along with a strong interaction potential leads to freezing of the dynamics of the second spin.} 
	\label{fig:2}
\end{figure*}

\begin{figure*}[!tbp]
	\centering

	\subfigure[]{\includegraphics[width=0.3\textwidth]{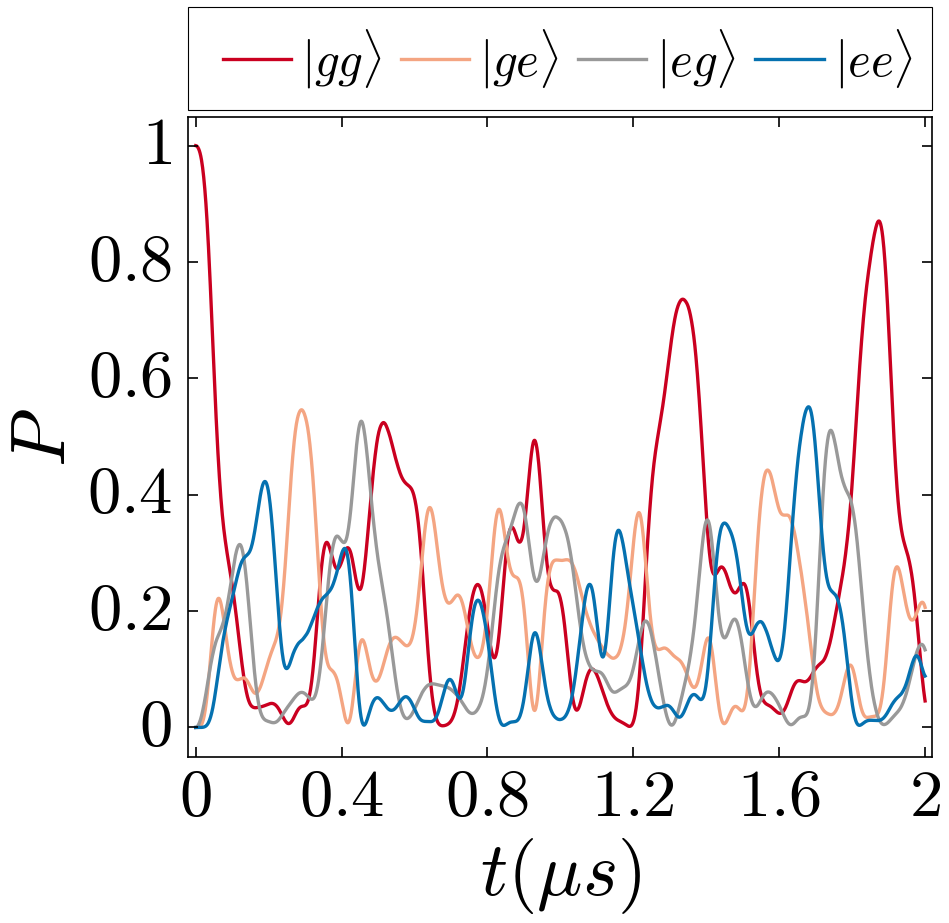}\label{3a}}
	\quad
	\subfigure[]{\includegraphics[width=0.3\textwidth]{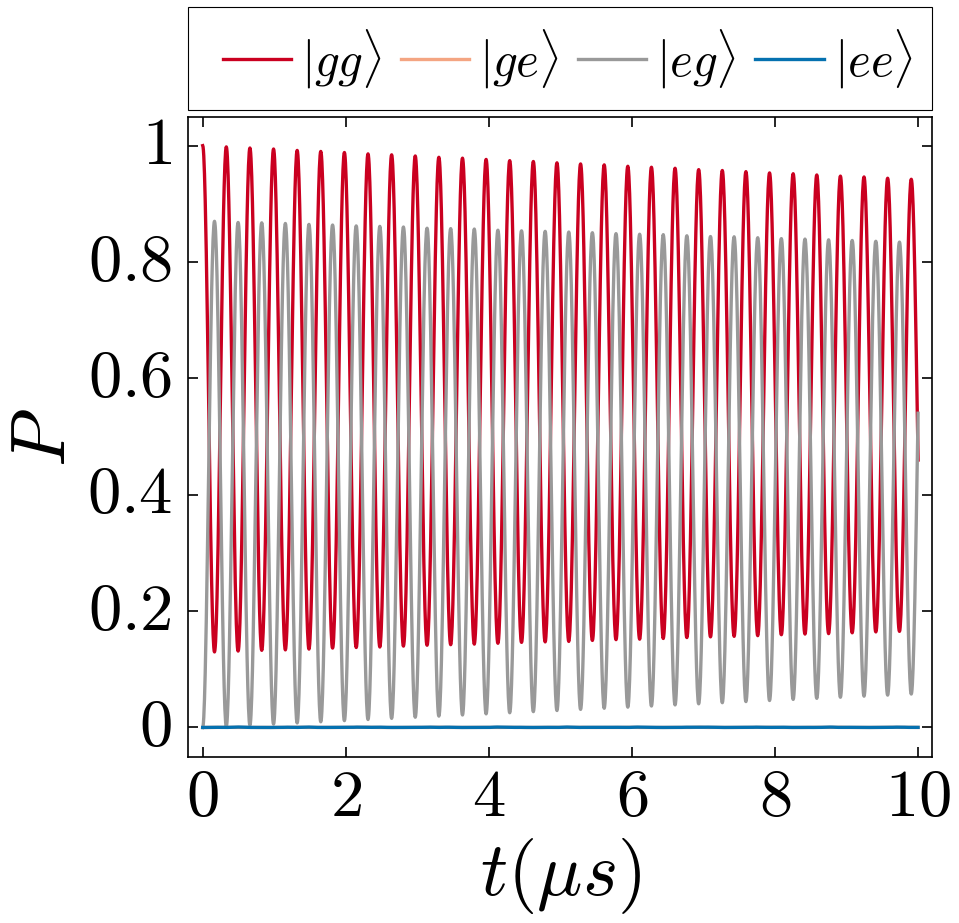}\label{3b}}
	\quad	
	\subfigure[]{\includegraphics[width=0.3\textwidth]{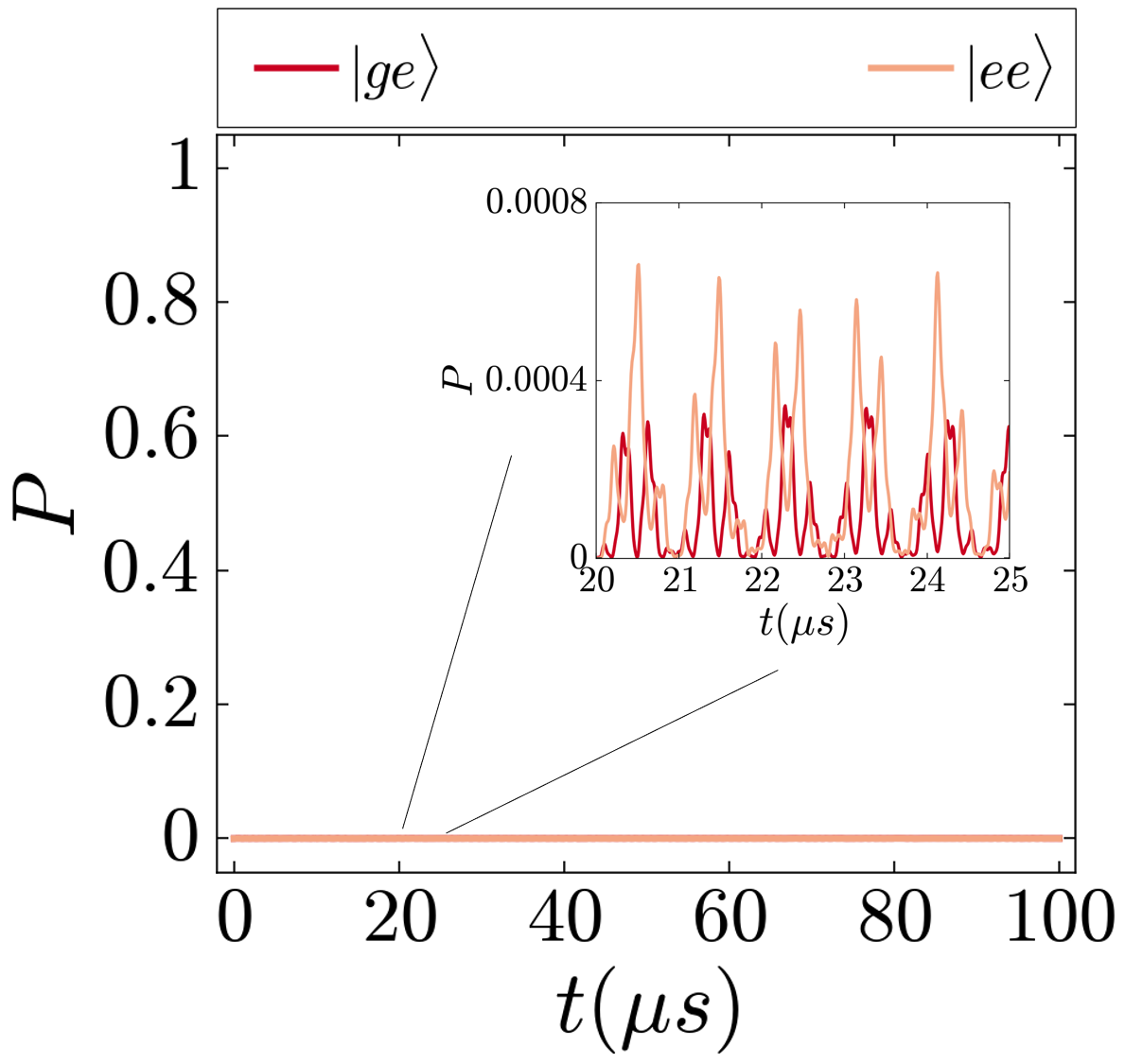}\label{3c}}
	\quad	
	
	\caption{State dynamics of the NV electron-nuclear spin system, hyperfine coupled with interaction potential  $A_{\parallel}=-2.16$MHz. (a) State dynamics for $\Omega_{MW}=\Omega_{RF}=4$MHz. Both electron and nuclear spin oscillate between their ground and excited states. (b) State dynamics for $\Omega_{MW}, A_{\parallel}>>\Omega_{RF}$  ($\Omega_{MW}=4$MHz, $\Omega_{RF}=40$kHz, $A_{\parallel}=-2.16$MHz). The system oscillates between the states $|gg\rangle$ and $|eg\rangle$, with the nuclear spin state frozen to the initial state $\ket{g}$. (c) A longer time evolution of the states $\ket{ge}$ and $\ket{ee}$  for the same choice of parameters as (b) to clearly depict interaction-induced nuclear spin freezing.}
	\label{fig:3}
\end{figure*}

\begin{figure*}[!tbp]
	\centering

	\subfigure[]{\includegraphics[width=0.35\textwidth]{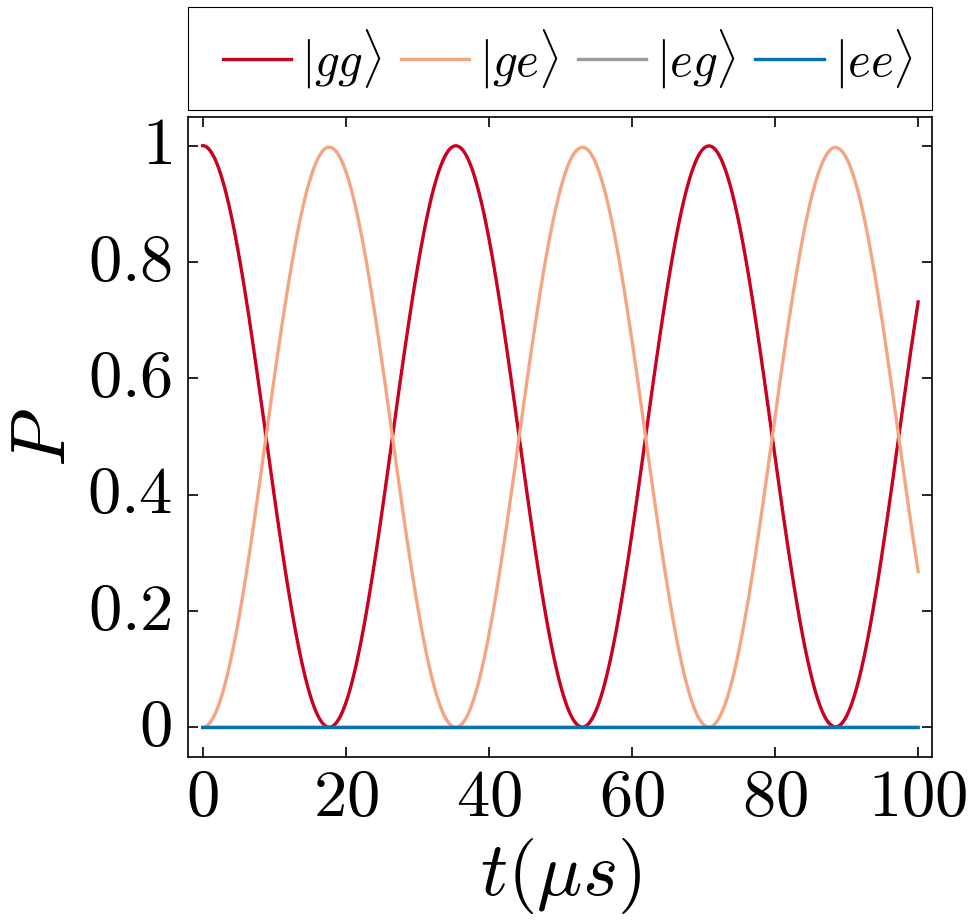}\label{4a}}
	\quad
	\subfigure[]{\includegraphics[width=0.35\textwidth]{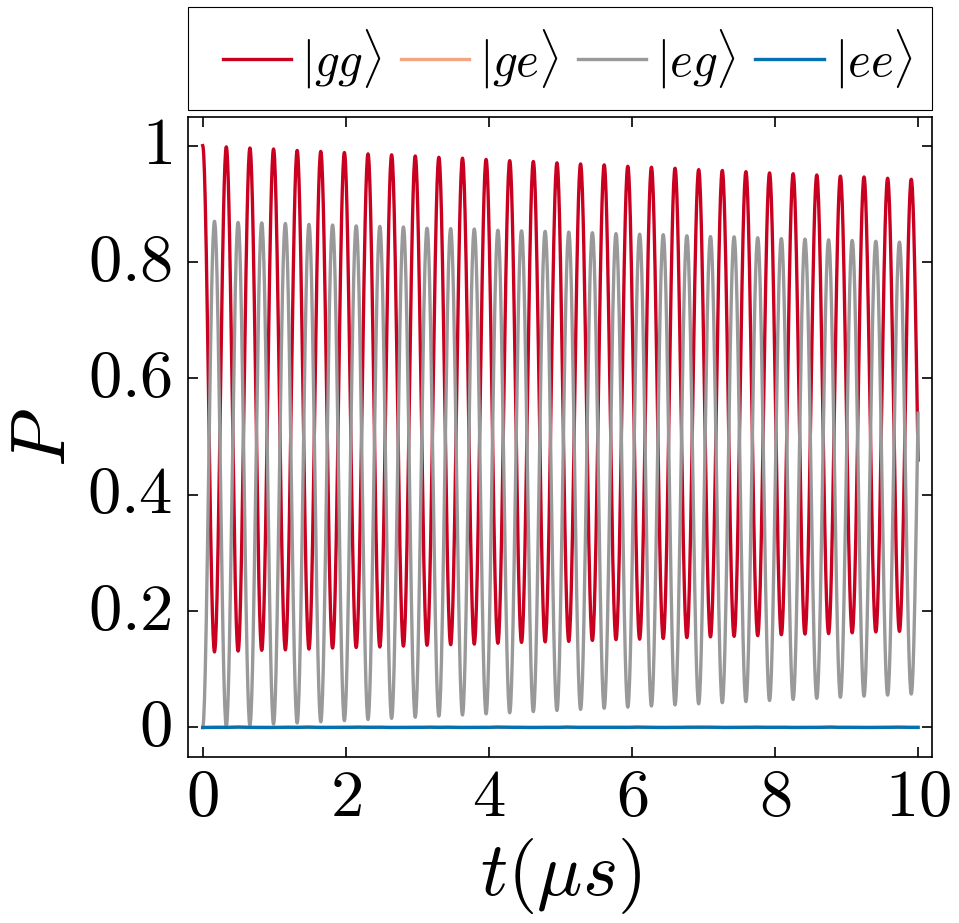}\label{4b}}
	\quad	
	\subfigure[]{\includegraphics[width=0.35\textwidth]{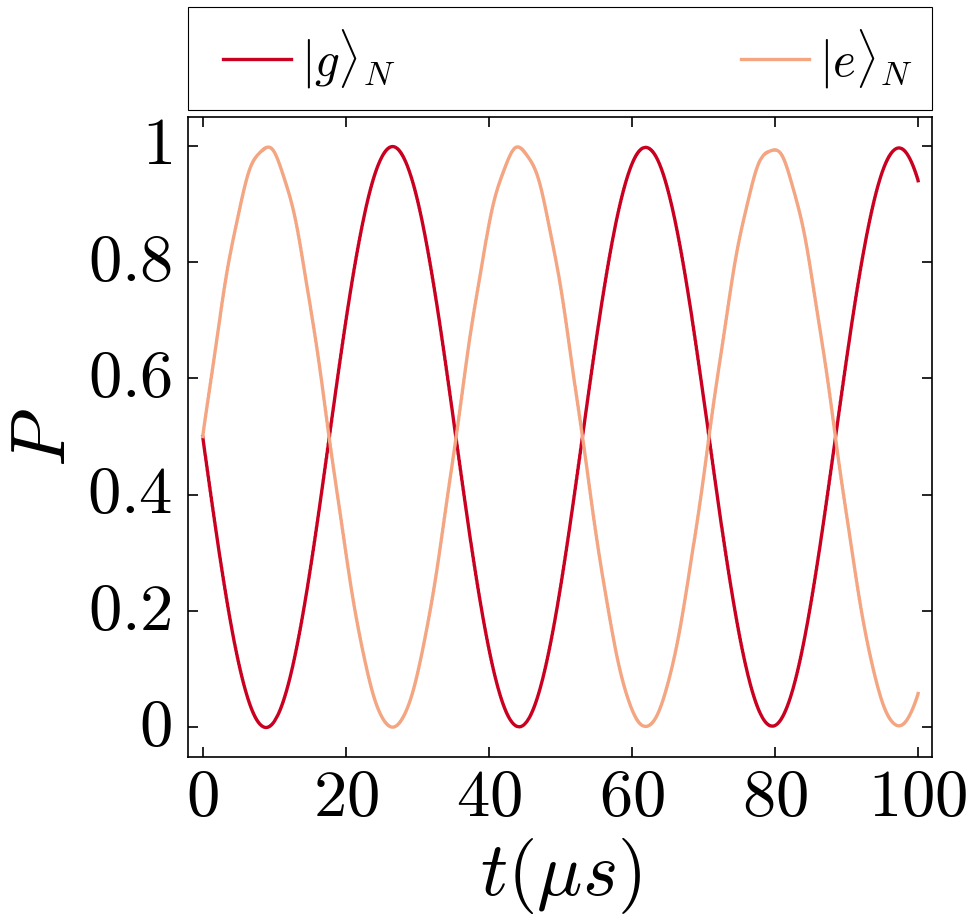}\label{4c}}
	\quad
	\subfigure[]{\includegraphics[width=0.35\textwidth]{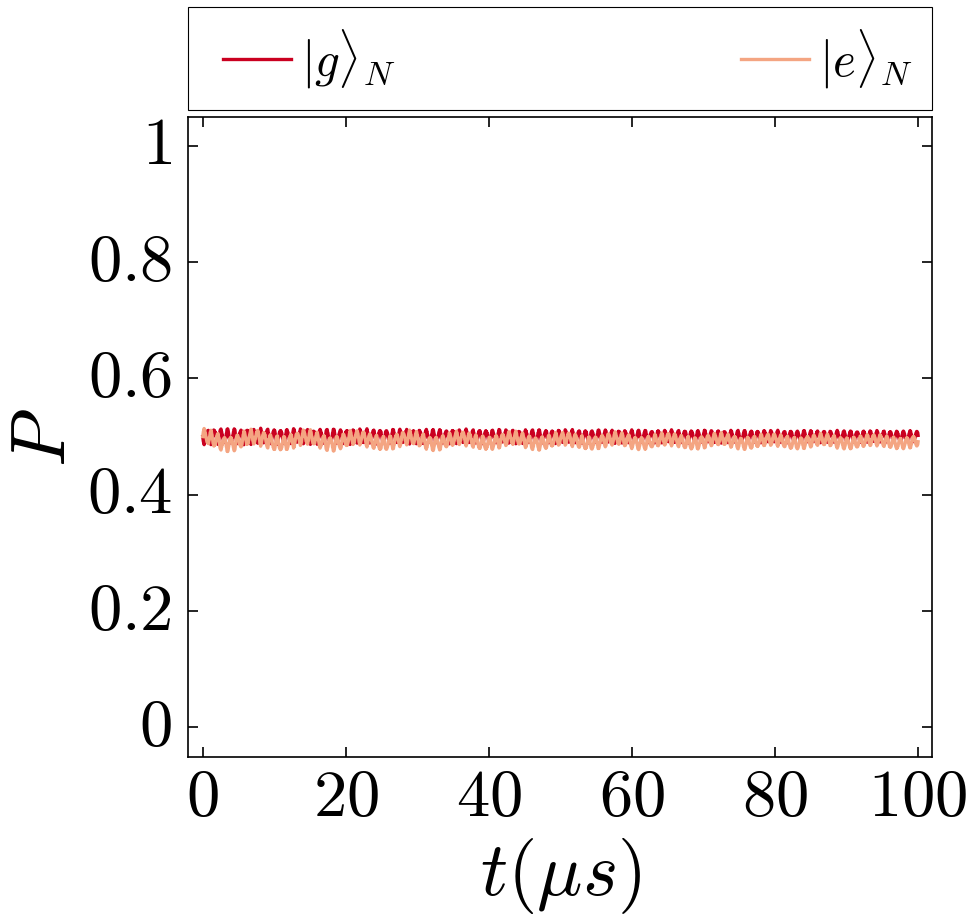}\label{4d}}
	\quad
	\caption{ Dynamics of the NV electron-nuclear state population (a) in the presence of a strong RF noise field resonant with the $\ket{gg} \rightarrow \ket{ge}$ transition. The state oscillates between $\ket{gg}$ and $\ket{ge}$, indicating that the nuclear spin oscillates between its ground and excited states. (b) in the presence of a strong resonant RF noise field and a decoupling MW field. The state oscillates between $\ket{gg}$ and $\ket{eg}$, indicating that the nuclear spin state stays in its initial state $\ket{g}$ and is decoupled from the RF noise field. (c) for an initial equal superposition state of the nuclear spin in the presence of a resonant RF field. The population oscillates between the ground and excited states of the nuclear spin (d) for an initial equal superposition state of the nuclear spin in the presence of a resonant RF field and a decoupling MW field. The population stays close to $0.5$ for the ground and excited states of the nuclear spin, indicating the nuclear spin state population stays close to the initial state populations.     }
	\label{fig:4}
\end{figure*}

In this work, we develop a nuclear spin freezing protocol for NV centers based on interaction-induced freezing of driven spin systems.  To explain the observed dynamics, we establish the connection between freezing dynamics in the physical NV context to that of an abstract model of two-interacting spins.
We then extend this to a more general demonstration of shielding the NV nuclear spin state from strong environmental noise which can potentially be relevant for quantum technologies.
 In particular, we consider broadband environmental noise models with Gaussian and uniform amplitude distributions over several frequency ranges.
We illustrate how the protocol can freeze the nuclear spin to any initial superposition state for time scales of the order of electron spin coherence time, thus effectively shielding the nuclear spin from the environmental noise. These are the main results of our work which are summarized in Figs.~\ref{fig:2},\ref{fig:3},\ref{fig:4},\ref{fig:5}. \\

We additionally report the time-evolution of quantum correlations between the NV electronic and nuclear spin state by evaluating the quantum discord \cite{Henderson_2001,PhysRevLett.88.017901,PhysRevA.77.042303} of the composite system. We find that the correlations are strongly suppressed in the frozen nuclear spin-state regime as one might already suspect from the previous discussions, confirming the ability of the protocol to effectively decouple the NV center's electron and nuclear spins. The quantum discord measure helps to check this expectation even for cases when the electron and nuclear spins are initially correlated in some fashion. The quantum discord results along with spin dynamics results will demonstrate that our freezing protocol effectively isolates the nuclear spin from  electromagnetic noises and applied fields as well as the NV center's electron spin. This nuclear spin freezing effect could potentially allow for nuclear spin-based quantum memories with longer storage times. \cite{doi:10.1126/science.1176496,doi:10.1126/science.1139831,Fuchs2011, PhysRevX.9.031045}. \\

\section{Setup and Methodology}\label{sec2}

 The NV center is a defect center in the diamond lattice with a substitutional Nitrogen atom and a vacancy on adjacent lattice sites. Figure \ref{1a} depicts the physical structure of an NV center in a diamond lattice. The NV center's ground state manifold is a spin-triplet ($S=1$) with electronic spin projection states $\ket{m_{S}=0,\pm 1}$. Coherent manipulations in the ground state manifold are performed by applying microwave field of appropriate frequency. The NV center also possesses an intrinsic \textsuperscript{14}N nuclear spin ($I=1$), which couples to the electronic spin via hyperfine coupling and splits each electronic level into three levels corresponding to the nuclear spin projection states $\ket{m_{I}=0,\pm 1}$. For a specific electron spin state, the nuclear spin transitions are performed using resonant RF fields. Fig. \ref{1c} depicts the energy level diagram of an NV center with the electron and nuclear spins interacting via hyperfine coupling. \\

The ground triplet state manifold of the NV center is governed by the Hamiltonian  \cite{DOHERTY20131}:

\begin{equation} \label{3}
\begin{aligned}
H_{0} =2\pi [D S_{z}^{2}+g_{e} \mu_{B} B_{z}  S_{z}+&A_{\perp}\left(S_{x} I_{x}+S_{y} I_{y}\right) +\\
&A_{\|} S_{z} I_{z} + Q I_{z}^{2}-g_{N} \mu_{N} B_{z}  I_{z}]
\end{aligned}
\end{equation}

Here $D = $ \SI{2.87}{\giga \hertz} is the zero-field splitting which separates the $\ket{0}$ and $\ket{\pm1}$ electronic states. On the application of a Zeeman field along the z-axis denoted by $B_z$, where z-axis is along the symmetry axis of the NV center, the $\ket{+1}$ and $\ket{-1}$ states split with the separation of $2g_{e}\mu_{B}B_{z}$. Here $g_{e}$ denotes the electronic $g$ factor and $\mu_B$ denotes the Bohr-magneton with $g_{e}\mu_{B}=2.802$ MHz/G. Hyperfine coupling to the \textsuperscript{14}N nuclear spin intrinsic to the NV center, further splits each of these NV electronic states to three states corresponding to the nuclear spin projections $m_{I}=-1,0,1$. For each electronic spin state, the nuclear spin state $\ket{m_{I}=0}$ is separated from the $\ket{m_{I}=\pm 1}$ states by nuclear quadrupole coupling $Q = -4.962$ MHz. The degeneracy of the $\ket{m_{I}=\pm 1}$ is lifted by the nuclear Zeeman splitting given by $g_{N}\mu_{N}B_{z}$, where $g_{N}$ denotes the nuclear $g$ factor and $\mu_N$ denotes the nuclear-magneton with $g_{N}\mu_{N}=0.308$ kHz/G. For the electron spin projections $m_{S}=-1,1$, the nuclear spin states $\ket{m_{I}=\pm 1}$ are further split by hyperfine coupling to the electronic spin. The hyperfine coupling is represented by the tensor $A$, which has longitudinal and transverse components $A_{\parallel}= -2.16 $ MHz and $A_{\perp}=-2.70$ MHz respectively. By confining the dynamics to the $\{\ket{0},\ket{-1}\}_{e}\otimes \{\ket{0},\ket{+1}\}_{N}$ subspace, the electron and the nuclear spins behave as an interacting two-qubit system. The Hamiltonian $H_0$, in this reduced subspace is then given by the simplified relation:

\begin{equation} \label{4}
\begin{aligned}
H_0= \pi\Biggl[&-\left(D-g_{e}\mu_{B}B_{z}-\frac{A}{2}\right) \sigma_z \otimes I + \\
&\left(Q+g_{N}\mu_{N}B_{z}-\frac{A}{2}\right) I \otimes \sigma_z+\frac{A}{2} \sigma_z \otimes \sigma_z\Biggr]
\end{aligned}
\end{equation}
 \\

  To observe interaction-induced freezing dynamics in NV centers, however, we need a resonant drive term in the Hamiltonian. This requires simultaneous driving of the NV electron spin and the intrinsic \textsuperscript{14}N nuclear spin. The electron and nuclear spin transitions can be driven by applying microwave and radio frequency(RF) fields, respectively. We confine the system dynamics to the $\{\ket{0},\ket{-1}\}_{e}\otimes \{\ket{0},\ket{+1}\}_{N}$ subspace by appropriately choosing frequencies of the drive fields. The electron spin is driven by a microwave field with a frequency mid-way between the transition frequencies for the electronic state transitions $\ket{0,0} \rightarrow \ket{-1,0}$ and $\ket{0,+1} \rightarrow \ket{-1,+1}$. Since both the transitions are equally detuned, they are driven with the same effective Rabi frequencies, and thus electronic transitions are driven irrespective of the nuclear spin state with the same Rabi frequency. The electronic drive Hamiltonian can be written as: 
\begin{equation} \label{5}
\begin{aligned}
H^{MW} = \Omega_{MW}\sin(2\pi f_{MW}t)S_{x}
\end{aligned}
\end{equation}

The nuclear spin state is driven using an RF field resonant with either of the transitions $\ket{0,0} \rightarrow \ket{0,+1}$ or $\ket{-1,0} \rightarrow \ket{-1,+1}$. From this, we can write the nuclear-spin drive Hamiltonian as:

\begin{equation} \label{6}
\begin{aligned}
H^{RF} = \Omega_{RF}\sin(2\pi f_{RF}t) I_{x}
\end{aligned}
\end{equation}

Thus in the presence of simultaneous nuclear and electron spin driving, the net Hamiltonian is given by:

\begin{equation} \label{7}
\begin{aligned}
H = H_{0}+H_{MW}+H_{RF}
\end{aligned}
\end{equation}

The dynamics of the states are simulated using the Lindblad master equation \cite{2020_lindblad}. Further, we introduce electron dephasing mechanism in the simulations with a characteristic $T_2$ time of $150 \mu$s. The nuclear spin dephases on a much longer time scale and can be neglected for the time scale of our simulations. The electron dephasing is taken into account by using a Lindblad operator:

\begin{equation} \label{8}
\begin{aligned}
L = \sqrt{1/T_{2}}S_{z}
\end{aligned}
\end{equation}

The quantum correlations in a multi-qubit system can be gauged by multiple metrics. One the most obvious choice for assessing the quantumness of the correlation is by calculating the entanglement entropy \cite{PhysRevA.54.3824,PhysRevA.53.2046}. However, entanglement entropy suffices as a good metric only if the state of the composite system is a pure state. We examine a more generic metric for measuring quantum correlations termed quantum discord since we are interested in quantum correlations at time scales equivalent to electron spin coherence durations \cite{Henderson_2001,PhysRevLett.88.017901,PhysRevA.77.042303}. Quantum discord acts as a more generic metric to gauge non-classical correlations in a system. The idea behind quantum discord originates from classical mutual information between two partitions in a bipartite system. Quantum discord can be quantified by evaluating the difference between the quantum analogue of two classically equivalent methods of measuring mutual information. \\

Mutual information in a bipartite system with partitions $A,B$ can be defined as:
\begin{equation}
I(\rho):=S\left(\rho^{A}\right)+S\left(\rho^{B}\right)-S(\rho)
\end{equation}

where $\rho_{A,B}$ are obtained taking a partial trace over the system density matrix $\rho$, and  $S(\rho)=-\text{Tr}\left(\rho\ln\rho\right)$ is the von Neumann entropy. \\

Alternatively, mutual information can be defined as:
\begin{equation}
Q_{A}(\rho)=S\left(\rho^{B}\right)-\sum_{k} p_{k} S\left(\rho^{B \mid k}\right)
\end{equation}
where $p_k$ is the probability of a certain outcome, $k$ on measuring the subsystem $A$ and $\rho^{B \mid k}$ is the state of subsystem $B$ based on the measured state of $A$. Classically these alternate definitions of mutual information are equivalent. However, since measurement in quantum systems is a fundamentally different process than in classical systems, these mutual information definitions lead to different values. The difference between these definitions depends on the basis ($\{E_{k}\}$) in which the subsystem $A$ is measured.
Quantum discord is defined as:
\begin{equation} \label{eq11}
D_{A}(\rho)=I(\rho)-\max_{\{E_{k}\}} Q_{A}(\rho)
\end{equation}

where $\max_{\{E_{k}\}} Q_{A}(\rho)$ gives a measure of the total classical correlations in the composite system. Subtracting classical correlations from the net correlation $I(\rho)$ quantifies the quantum correlations in the system and is known as the quantum discord of the system. There is no closed-form analytic expression for quantum discord for a general two-qubit quantum state. Thus, calculating quantum discord for a general density matrix involves brute force maximization of $Q_{A}(\rho)$ over a large number of orthonormal measurement bases. We obtain $Q_{A}(\rho)$ values by measuring subsystem $A$ in the orthonormal basis: $\left\{|u\rangle=\cos \theta|0\rangle+e^{i \phi} \sin \theta|1\rangle,|v\rangle=\sin \theta|0\rangle-e^{i \phi} \cos \theta|1\rangle\right\}$ for a large number of values of $\theta$ and $\phi$. The maximum value of $Q_{A}(\rho)$ obtained from the optimization provides us a measure of the classical correlations in the system. \\

Finally, we draw a natural parallel to a general setup of two interacting spins driven by resonant fields to highlight the importance of different parameters in an interaction-induced freezing experiment and to explain the results observed for NV centers. The spins are modeled as two-level systems with ground state $\ket{g}$ and excited state $\ket{e}$. The spins interact via a potential $V_{0}$. Fig. \ref{1b} depicts the energy level diagram of two such interacting spins. The Hamiltonian for this system can thus be written as:
 
 \begin{equation}  \label{1}
H=-\sum_{i=1}^{2} \Delta_{i} \sigma_{e e}^{i}+\sum_{i=1}^{2} \frac{\Omega_{i}}{2} \sigma_{x}^{i}+V_{0} \sigma_{e e}^{1} \sigma_{e e}^{2}
\end{equation}

where $\Delta_{i}$ and $\Omega_{i}$ denote the detuning and the Rabi frequency of the drive field applied on the $i^{\text{th}}$ spin, and $V_0$ represents the interaction potential between the two spins. $\sigma_{x}$ represents the Pauli-sigma $x$ operator whereas $\sigma_{e e}$ is defined as $\sigma_{e e}=\ket{e} \bra{e}$ . \\

Assuming resonant drive fields, that is, $\Delta_{i}=0$, we get: 

\begin{equation}  \label{2}
H=\sum_{i=1}^{2} \frac{\Omega_{i}}{2} \sigma_{x}^{i}+\frac{V_{0}}{4} (\mathbb{I}-\sigma_{z}^{1})(\mathbb{I}-\sigma_{z}^{2})
\end{equation}

where $\sigma_{e e}$ is expressed in terms of the Pauli-sigma operators as $(\mathbb{I}-\sigma_{z})/2$. \\

In the following section we present the results obtained from numerical simulations of time evolution of state population and quantum correlations using the methods discussed above. We start by discussing the results for interaction-induced freezing in the general setup of two interacting spins to highlight the parameter regimes where the freezing protocol works. We then present results  on interaction-induced freezing for the specific parameter values relevant to NV centers (Eq.~\eqref{7} and the preceding discussion), by evolving the NV center under \eqref{7} and extend these results to and demonstrate the decoupling of the NV center's nuclear spin from strong noise fields. Finally, we assess the decoupling of the NV center's electron and nuclear spins by evaluating the evolution of quantum discord using Eq.~\eqref{eq11}

\section{Results and Discussion}\label{sec4}

Fig. \ref{fig:2} depicts the oscillations in state populations with respect to time when two spins, interacting with an interaction potential $V_{0}$, are continuously and independently driven  on resonance by different drive fields with Rabi frequencies $\Omega_1$ and $\Omega_2$ respectively. The spins can each occupy their ground or excited states denoted by $\ket{g}$ and $\ket{e}$, respectively. The collective state of the two-spin system is a superposition in the basis states given by $\{\ket{g}, \ket{e} \} \otimes \{\ket{g}, \ket{e} \}$. First, both spins are initialized to their ground states, which initializes the composite system to the state $\ket{gg}$. Fig. \ref{2a} depicts the state dynamics when the interaction potential $V \approx 0$. Experimentally, this corresponds to the scenario where the spins are placed far from each other. On turning on the drive fields such that $\Omega_{1}=\Omega_{2}=2$MHz, we observe that the system state coherently oscillates between the states $\ket{gg}$ and $\ket{ee}$, i.e., between their ground and doubly excited states, as expected for two independent spins driven on-resonance. To observe the effect of interaction potential on the dynamics, we bring the spins close to each other such that $V_{0}$ increases to $2$MHz. Keeping the Rabi frequencies for the two drive fields fixed at  $\Omega_{1}=\Omega_{2}=2$MHz, we observe the population oscillates between all four basis states as depicted in Fig. \ref{2b}, and the excited states for both spins get populated during the dynamics. Next, we move the spins away from each other to reduce the interaction potential back to $V \approx 0$. To observe the effect of a high bias in the Rabi frequencies of the applied drive fields, the Rabi frequency of the second spin is decreased, such that $\Omega_{1}>>\Omega_{2}$  $(\Omega_{1}=2\text{MHz}, \Omega_{2}=100\text{kHz})$. As shown in \ref{2c}, in a shorter timescale, the system shows fast oscillations between the states $\ket{gg}$ and $\ket{eg}$ due to the higher Rabi frequency of the drive field on the first qubit. The states $\ket{ge}$ and $\ket{ee}$, however, eventually get populated, and we observe oscillations between all four states of the system at a longer timescale due to the lower Rabi frequency of the drive field on the second qubit. We have individually observed the effect of introducing a high interaction potential and a high bias in the applied fields in Figs. \ref{2b} and Fig. \ref{2c} respectively, where both the spins occupy their excited states during the dynamics, and thus none of the spins have a fixed state. On simultaneously applying these two conditions, i.e., a strong interaction potential ($V_{0}>>\Omega_{2}$) and a high bias in the Rabi frequencies ($\Omega_{1}>>\Omega_{2}$) : $\Omega_{1}=2$ MHz, $\Omega_{2}=100$ kHz, $V_{0}=2$ MHz, we observe that the dynamics of the spin driven with a lower Rabi frequency drive field, i.e., the second spin, completely freezes, demonstrating the interaction-induced freezing phenomenon. This can be seen in Fig. \ref{2d}, where the state of the system oscillates between $\ket{gg}$ and $\ket{eg}$ and the population of $\ket{ge}$ and $\ket{ee}$ stay close to zero. This indicates that the state of the second spin is frozen to its initial state $\ket{g}$ and is unaffected by the applied drive fields. Therefore, strong interaction potential and a high bias in the applied Rabi frequencies are essential to observe interaction-induced freezing. \\

\begin{figure*}[!tbp]
	\centering

	\subfigure[]{\includegraphics[width=0.35\textwidth]{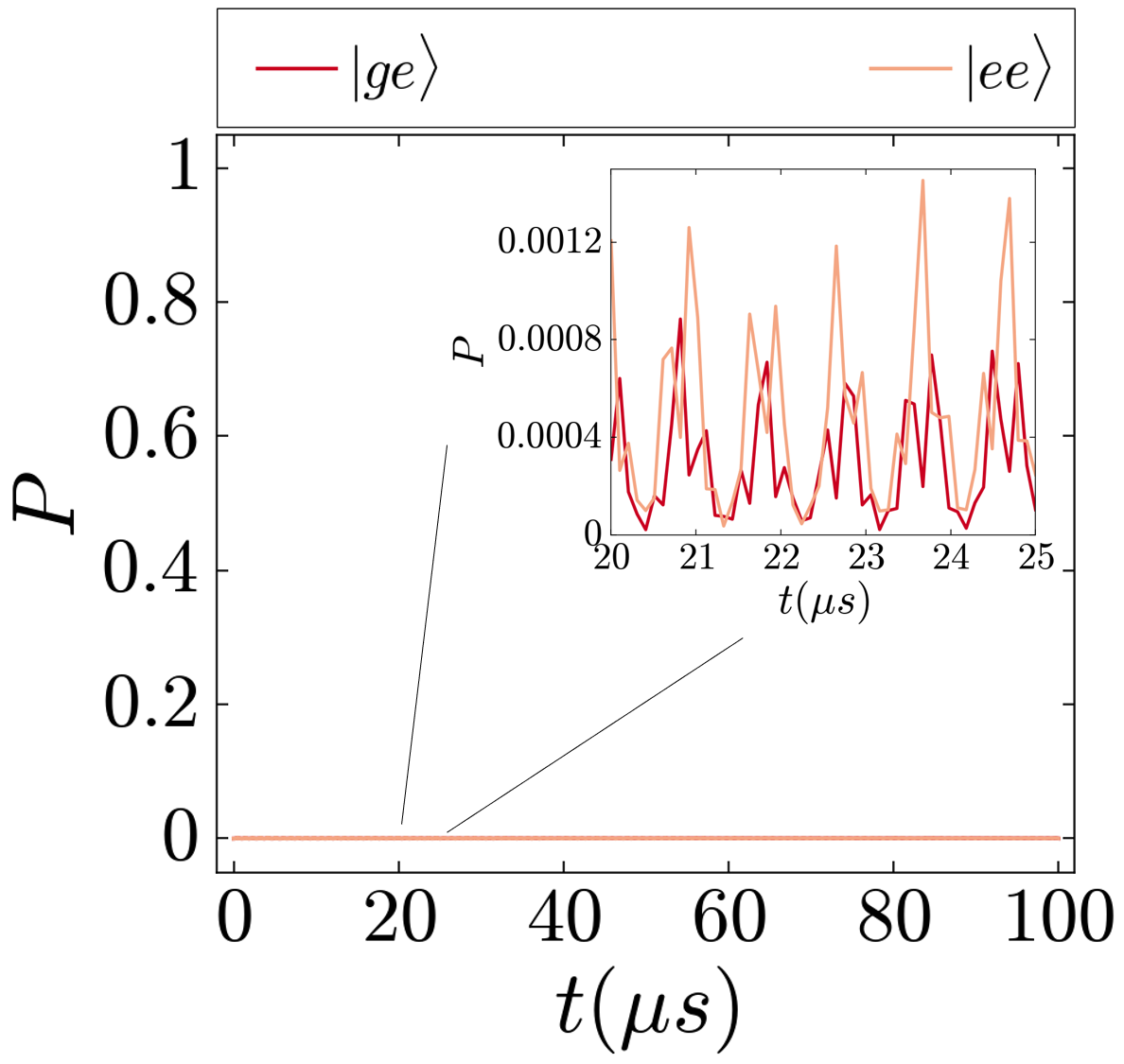}\label{5a}}
	\quad
	\subfigure[]{\includegraphics[width=0.35\textwidth]{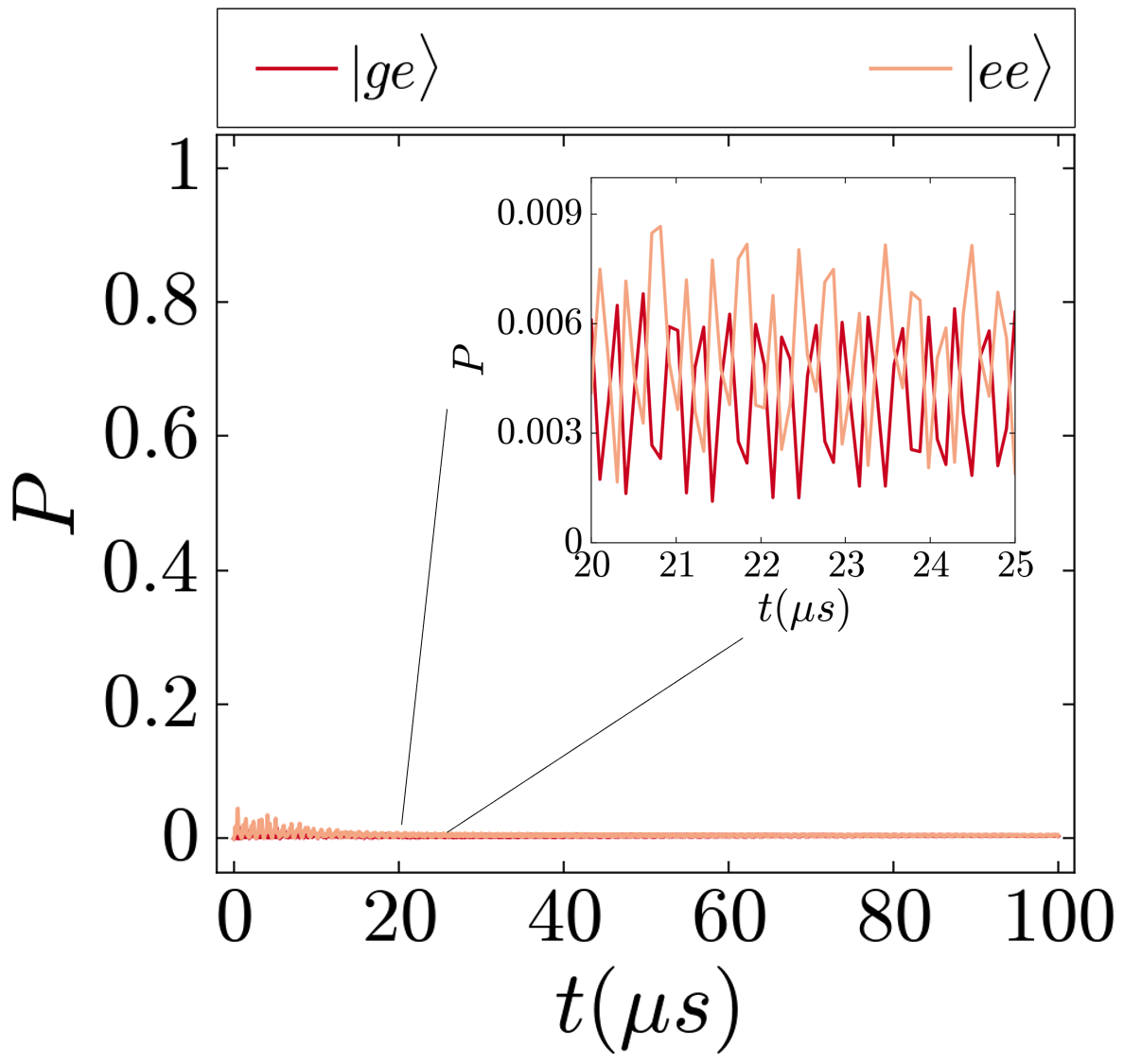}\label{5b}}
	\quad	
	\subfigure[]{\includegraphics[width=0.35\textwidth]{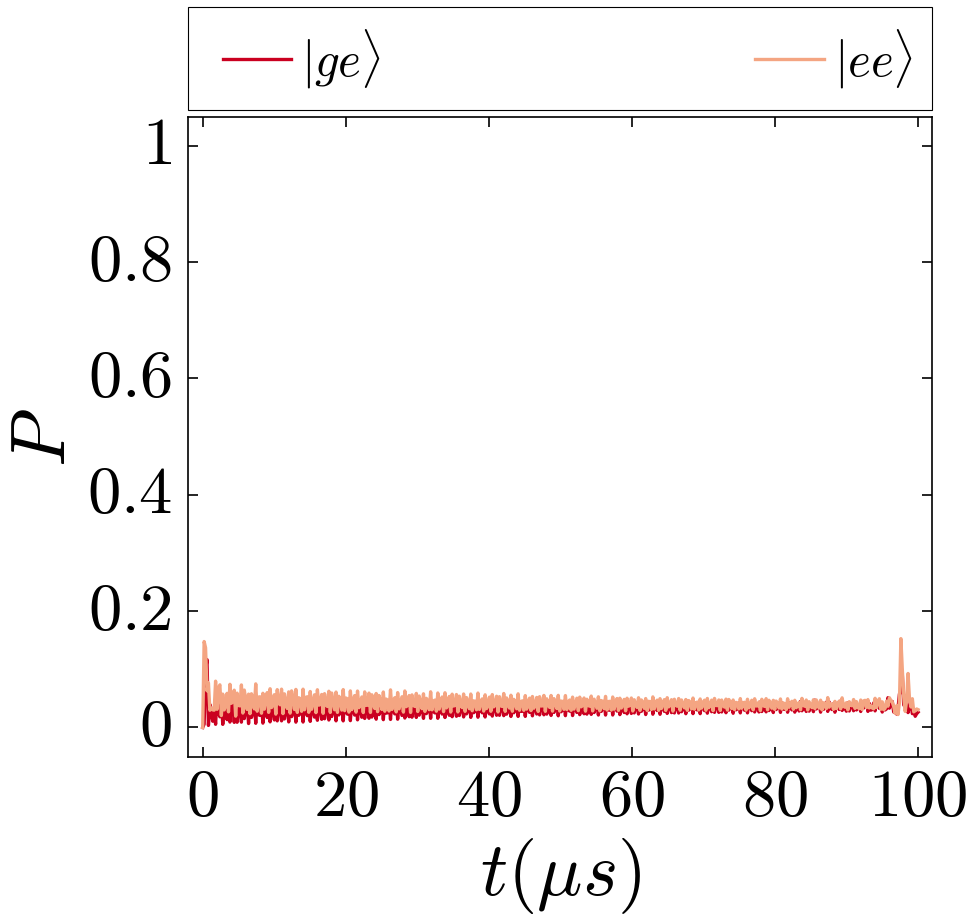}\label{5c}}
	\quad
	\subfigure[]{\includegraphics[width=0.35\textwidth]{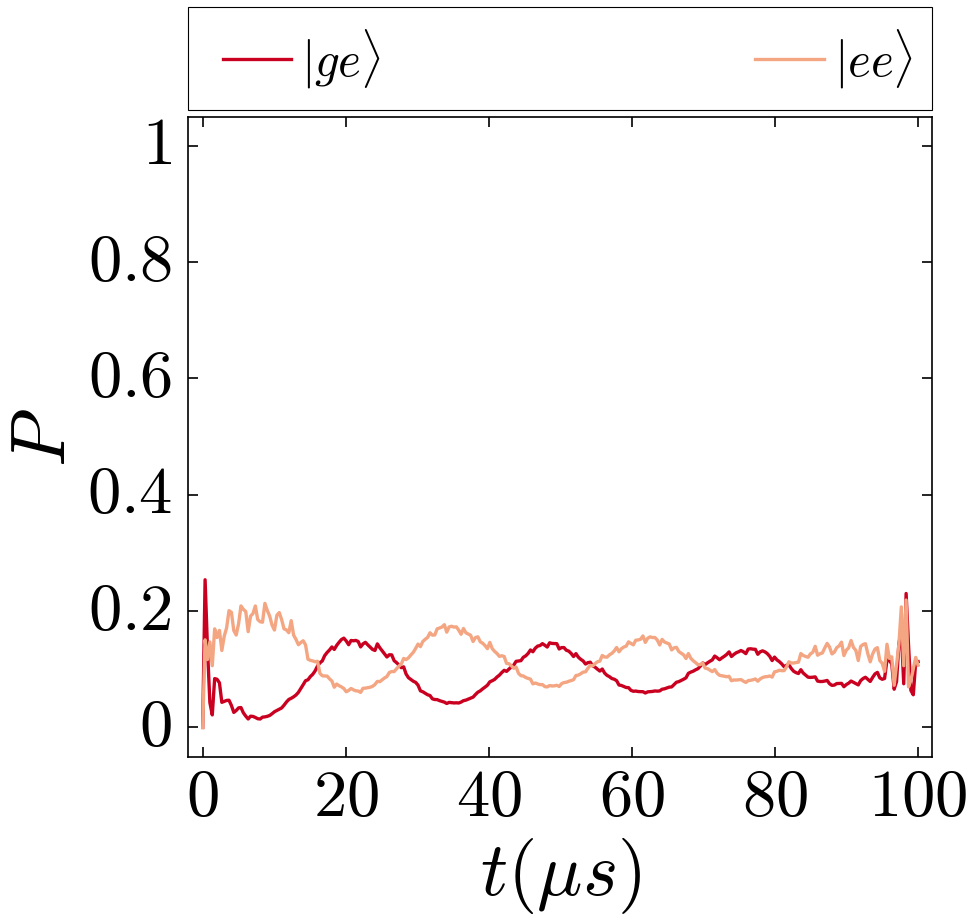}\label{5d}}
	\quad	
	\caption{ State dynamics of the electron-nuclear spin system in the presence of the strong microwave field ($\Omega_{MW}= 4$MHz) and (a) a Gaussian frequency profile RF noise field with amplitude=$30$KHz and $\sigma=10$KHz. (b) a Gaussian frequency profile RF noise field with amplitude=$30$KHz and $\sigma=100$kHz. (c) an RF noise field with a uniform amplitude of $10$KHz over the frequency range $(\omega_{RF}-0.5\text{MHz})$ to $(\omega_{RF}+0.5\text{MHz})$ (d) an RF noise field with a uniform amplitude of $20$KHz over the frequency range $(\omega_{RF}-0.5\text{MHz})$ to $(\omega_{RF}+0.5\text{MHz})$.  For general environmental noise fields, in the presence of the strong MW field, the nuclear spin stays pinned at its initial state $|g\rangle_{N}$ and thus, the population of states $|ge\rangle$ and $|ee\rangle$ remain close to zero.} 
	\label{fig:5}
\end{figure*}

Fig. \ref{fig:3} shows interaction-induced freezing now in the NV context with physically relevant parameter values, by simultaneously driving the NV electron and intrinsic nuclear spin and by evolving the system under the Hamiltonian described in equations \eqref{3}-\eqref{7}. As pointed out earlier, under the chosen drive fields, the NV electron spin is a superposition of the basis states $\{\ket{0}, \ket{-1}\}_{e}$ which, for simplicity, we denote by the notation $\{\ket{g}, \ket{e}\}_{e}$. Similarly, the intrinsic nuclear spin is a superposition of the basis states $\{\ket{+1}, \ket{0}\}_{N}$, which we denote by $\{\ket{g}, \ket{e}\}_{N}$. Thus the state of two-qubit electron-nuclear composite system is a superposition of the basis states $\{\ket{g}, \ket{e} \} \otimes \{\ket{g}, \ket{e} \}$. We start by initializing the state of the system to $\ket{gg}$ where both the nuclear and electronic spins are in their ground state.  Unlike the general model for interacting spins, where the interaction potential was controllable based on the distance between the spins, the interaction potential between the NV center's electron and nuclear spins is fixed at  $A_{\parallel}=-2.16$ MHz because of the hyperfine coupling between the electron and nuclear spins. Fig. \ref{3a} depicts the state dynamics for a scenario similar to Fig. \ref{2b}, where the Rabi frequencies for the applied drive fields on the electron and nuclear spin are equal ($\Omega_{MW} = \Omega_{RF} = 4$ MHz) and the spins interact via hyperfine coupling with $A_{\parallel}=-2.16$ MHz.  Similar to Fig. \ref{2b}, we observe that the state population oscillates between all four basis states, which implies that both the electron and the nuclear spin states oscillate between their ground and excited states. However, on fixing  $\Omega_{MW}$ to $4$ MHz and decreasing $\Omega_{RF}$ to $100$ kHz, with the spins still interacting via hyperfine coupling $A_{\parallel}=-2.16$ MHz, we observe the state dynamics of the nuclear spin freezes. This is shown in Fig. \ref{3b}, where the state population oscillates between $\ket{gg}$ and $\ket{eg}$. The choice of parameters is similar to Fig. \ref{2d}, where we have a high bias in the Rabi frequencies of the applied fields and a strong interaction potential.  The population of the states $\ket{ge}$ and $\ket{ee}$ stays close to zero, implying that the nuclear spin state remains fixed at the initial state $\ket{g}$~\cite{footnote_population}. 
 This is thus an analogous effect to the interaction-induced freezing depicted in Fig. \ref{2d}. Fig. \ref{3c} demonstrates the long-time behavior of the states for the same choice of parameters, thus showing that the nuclear spin state remains fixed at the initial state $\ket{g}$ on timescales much longer than $1/\Omega_{RF}$. We can observe from the preceding fingerprints that obtaining interaction-induced freezing requires a substantial interaction potential and a large disparity in the applied Rabi frequencies. \\

\begin{figure*}[!tbp]
	\centering

	\subfigure[]{\includegraphics[width=0.43\textwidth]{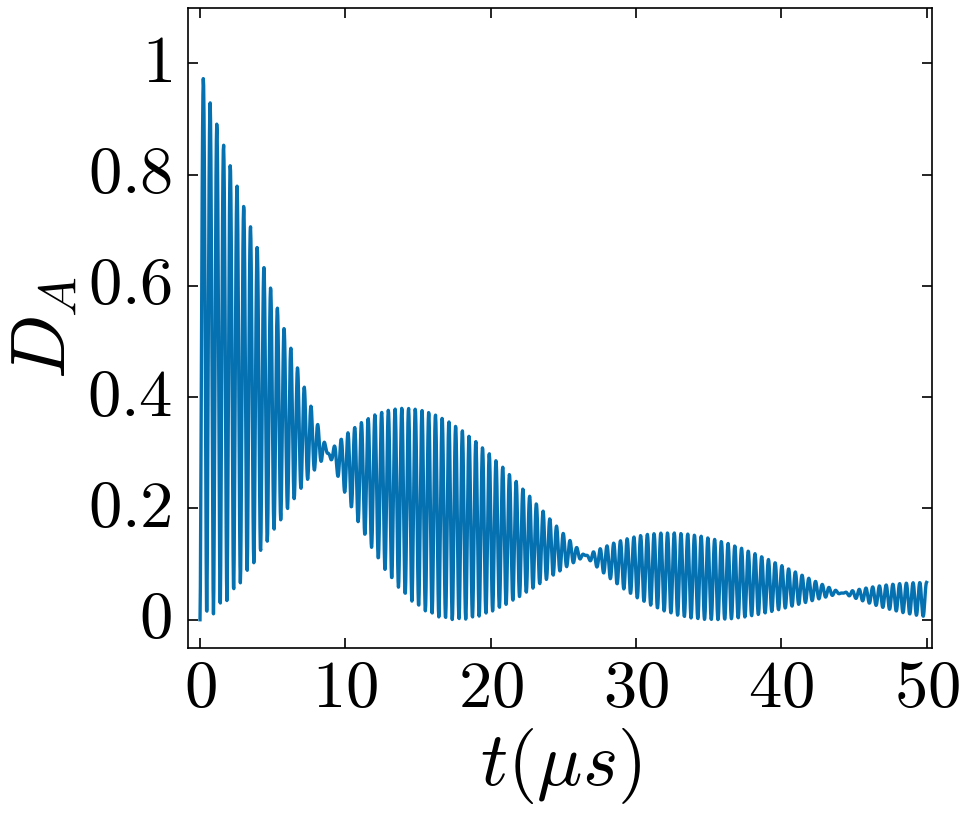}\label{6a}}
	\quad
	\subfigure[]{\includegraphics[width=0.43\textwidth]{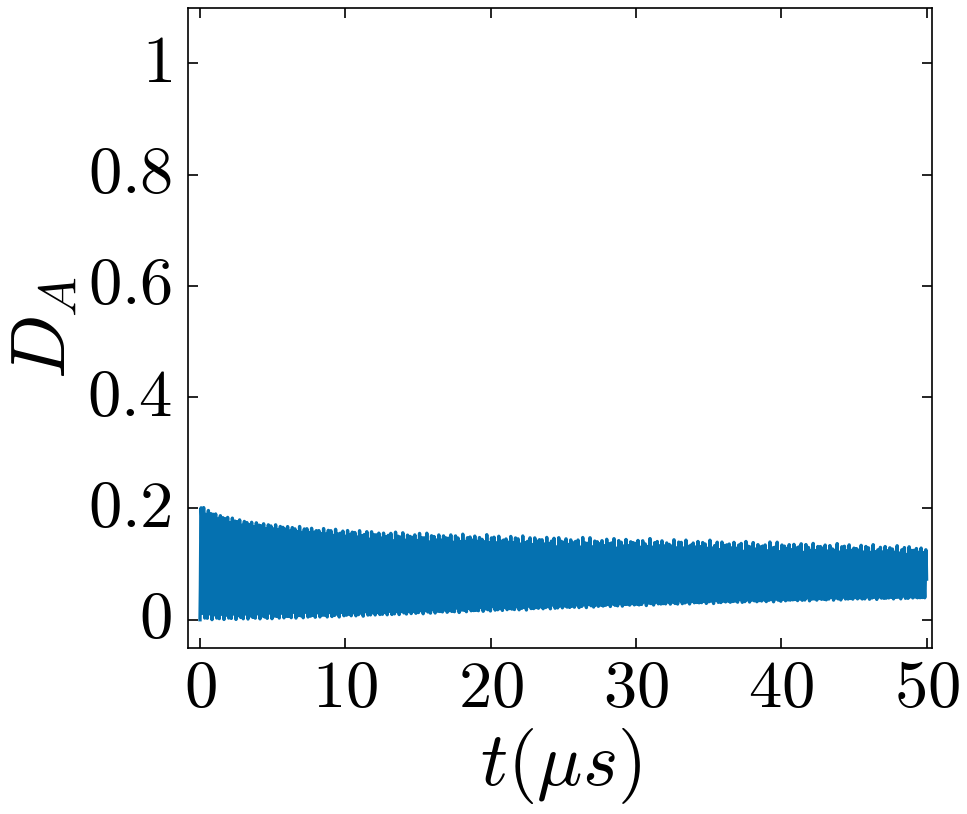}\label{6b}}
	\quad	

	\caption{ Evolution of quantum discord of the NV electron-nuclear spin system (a) in the presence of just the RF noise field ($\Omega_{RF}= 40$kHz) (b) in the presence of the RF noise field ($\Omega_{RF}= 40$kHz) and a decoupling MW field ($\Omega_{MW}= 4$MHz).  In the presence of just the RF noise field, the electron and nuclear spins get significantly correlated during the dynamics. In contrast, these correlations are suppressed on introducing the decoupling MW field. } 
	\label{fig:6}
\end{figure*}

Next, as shown in Fig. \ref{fig:4}, the study is extended to demonstrate that interaction-induced freezing protocol can effectively decouple the nuclear spin from strong external noise fields. The external noise fields are modeled by RF fields resonant with the $\ket{gg} \rightarrow \ket{ge}$ transition, with a Rabi frequency of $40$ kHz. First, as shown in Fig. \ref{4a}, for an initial system state of $|gg\rangle$, in the presence of an RF noise field without applying any MW field, we observe that the state oscillates between $|gg\rangle$ and $|ge\rangle$. This shows the presence of the RF noise field causes nuclear spin oscillations. However, on turning on a strong MW field with a frequency midway between the $|gg\rangle \rightarrow |eg\rangle$ and $|ge\rangle \rightarrow |ee\rangle$ transitions and a Rabi frequency much greater than the Rabi frequency of the RF noise field ($\Omega_{MW}= 4$MHz, $\Omega_{RF}= 40$ kHz), as shown in Fig. \ref{4b}, the nuclear spin state stays frozen at its initial state $|g\rangle$, and the electron spin oscillates between its ground and excited states indicating the decoupling of the nuclear spin from the RF noise field due to the applied MW field.  We next investigate the method's robustness to the nuclear spin's initial state. As shown in Fig. \ref{4c}, for an initial equal superposition nuclear spin state $(|g\rangle + |e\rangle)/\sqrt(2)$, in the presence of just the resonant RF noise field, the nuclear spin state oscillates between it's ground and excited states. In the presence of the strong MW field, however, the state population for the ground and excited-state stays pinned around $0.5$, depicting the freezing of the nuclear spin state to the initial equal superposition state, as shown in Fig. \ref{4d}. \\

Fig.~\ref{fig:4} showed the freezing of nuclear spin state in the presence of a single-frequency RF noise field that was chosen to be resonant with the nuclear spin transition. Fig.~\ref{fig:5} depicts the system state dynamics now in the presence of broadband RF noise fields, for an initial state $|gg\rangle$ demonstrating the robustness of the protocol to a generalized environmental noise that is likely to be present under realistic experimental conditions. In Figs. \ref{5a}-\ref{5b}, we consider an RF noise field with a Gaussian frequency profile of the form $A(x)=A_{0}\exp{-(x-w_{RF})^{2}/(2\sigma^2)}$, centered around the nuclear spin transition frequency.  For $\sigma=10$ kHz, and $A_{0}=30$kHz, with a strong MW field with $\Omega_{MW}= 4$MHz, we observe that the population of states $|ge\rangle$ and $|ee\rangle$ stays pinned at a near-zero value throughout the dynamics, as shown in Fig. \ref{5a}. Keeping $A_0$ fixed and increasing the width of the Gaussian profile to $100$ kHz, we still observe that the nuclear spin state is efficiently frozen to its initial state $|g\rangle$, as shown in Fig. \ref{5b}. Next, in Figs. \ref{5c}-\ref{5d}, we consider a noise field which has a uniform amplitude $K$ over a frequency range of $(w_{RF}-0.5\text{MHz})$ to $(w_{RF}+0.5\text{MHz})$. For $K$ values $<10 $KHz, the nuclear spin state stays close to $|g\rangle$, as depicted in Fig. \ref{5c}, where $K=10$KHz. Increasing $K$ further to $20$KHz leads to perceptible oscillations in the nuclear spin state, as shown in Fig. \ref{5d}. We thus infer that the method stays substantially robust for general noise fields for a wide range of frequency values and could be a potentially useful method to decouple the nuclear spin from environmental fields.\\

Finally, we discuss the dynamics of quantum correlations of the electron-nuclear system. We use the quantum discord ($D_A$), calculated using \eqref{eq11}, as a measure of quantum correlations. Starting with an equal superposition of the electron-nuclear basis states $\ket{gg}, \ket{ge}, \ket{eg}, \ket{ee}$, we evolve the system under the RF and MW fields and evaluate $D_A$ at each instance. In the presence of a single frequency RF noise field resonant with the $|gg\rangle \rightarrow |ge\rangle$ transition ($\Omega_{RF}= 40$kHz) and in the absence of a strong MW field, we observe that the quantum discord performs oscillations, attaining a maximum value of $\approx 0.5$, as shown in Fig. \ref{6a}. This indicates that evolution under the noise field substantially correlates the electron and nuclear spins. These correlations decay on the time scale of the electron spin coherence time. On turning on a strong  MW field however ($\Omega_{MW}= 4$MHz), these correlations are diminished, as shown in Fig. \ref{6b}, where $D_A$ stays close to $0$ throughout the dynamics. The nuclear spin freezing protocol can therefore effectively decouple the electron and nuclear spins. Thus, we have demonstrated that the advantage of applying a decoupling MW field is two-fold: decoupling the nuclear spin from noise fields and decoupling it from the electron spin and thus practically isolating it from it's environment. \\

\section{Conclusion}\label{sec13}
 To summarize, we presented interaction-induced freezing phenomenon in an NV center platform. Through thorough numerical simulations, we showed that the state dynamics of the NV center's intrinsic \textsuperscript{14}N nuclear spin freezes when the NV center's electronic and intrinsic nuclear spins, interacting via hyperfine-coupling, are simultaneously driven using microwave and RF fields with significantly different Rabi frequencies. We also numerically simulated the interaction-induced freezing phenomenon for a general system of two interacting spins by driving them using fields with highly different Rabi frequencies to highlight the important parameter regimes for interaction-induced freezing and compared these signatures to the corresponding NV center signatures.
We furthermore demonstrated the shielding of the NV center's nuclear spin from strong RF noise fields by applying a decoupling MW field. We note here that an earlier work~\cite{PhysRevLett.113.020506} which had focused on decoupling the nuclear spin from the drive fields has a similar flavor to the protocol given here, however we have checked for more general noise profiles in our work. We showed how the nuclear spin state population is preserved in presence of these noise fields for any general initial state and presented the method's robustness against general forms of noise fields. We evaluated the quantum correlations between the NV center's electron spin and nuclear spin states using quantum discord and observed significant suppression of quantum correlations in the decoupled or frozen nuclear spin regime. This gives a clear signature of decoupling of the nuclear spin from the electron spin. Thus, in this work, we demonstrated a protocol to isolate the NV center's nuclear spin from external fields as well as the NV electronic spin. Since this protocol effectively isolates the nuclear spin, it can have significant implications in extending the lifetimes of nuclear spin-based quantum memories.

\begin{acknowledgements}
K.S. acknowledges financial support from IIT Bombay seed grant number 17IRCCSG009, DST Inspire Faculty Fellowship-DST/INSPIRE/04/2016/002284 and DST Quest Grant DST/ICPS/QuST/Theme-2/2019/Q-58. 
S.P. acknowledges financial support from SERB-DST, India via grant no. SRG/2019/001419, and in the final stages of writing by grant no. CRG/2021/003024.
The authors acknowledge useful discussions with 
H. S. Dhar, A. Mahajan, and B. Muralidharan.
 
\end{acknowledgements}


\begin{thebibliography}{32}
\bibitem [{foo()}]{footnote_population}%
  \BibitemOpen
  \href@noop {} {}\bibinfo {note} {A slight difference in Figs. 2(d) and 3(b)
  is that the population of states $|gg\rangle$ and $|eg\rangle$ in Fig. 2(d)
  oscillates between $1$ and $0$, whereas in 3(b), the population of doesn't
  reach $0$. This is because we are using just one MW field to drive both the
  transitions $\{|0,0\rangle \rightarrow |-1,0\rangle\}$ and $\{|0,+1\rangle
  \rightarrow |-1,+1\rangle\}$. The applied MW has frequency midway between
  these two transitions to ensure both transitions are driven with the same
  Rabi frequency. Using a detuned Rabi drive leads to a drop in the change of
  population, and there can't be a perfect population transfer between
  $\{|0,0\rangle, |-1,0\rangle\}$ and $\{|0,+1\rangle,|-1,+1\rangle\}$. The
  imperfect driving due to a detuned drive field, however, still gives the
  interaction-induced freezing signature, which was the main motive of this
  protocol.}\BibitemShut {Stop}%
\end{thebibliography}
\end{document}